\documentclass[apj]{emulateapj}

\usepackage{apjfonts}
\usepackage{amssymb, amsmath}
\newcommand\degree{\degr}
\newcommand\degrees\degree
\newcommand\vs{\em vs.}

\DeclareSymbolFont{UPM}{U}{eur}{m}{n}
\DeclareMathSymbol{\umu}{0}{UPM}{"16}
\let\oldumu=\umu
\renewcommand\umu{\ifmmode\oldumu\else\math{\oldumu}\fi}
\newcommand\micro{\umu}
\renewcommand\micron{\micro m}
\newcommand\microns \micron

\renewcommand\arcsec[0]{$^{\prime\prime}$}

\let\oldsim=\sim
\renewcommand\sim{\ifmmode\oldsim\else\math{\oldsim}\fi}
\let\oldpm=\pm
\renewcommand\pm{\ifmmode\oldpm\else\math{\oldpm}\fi}
\newcommand\by{\ifmmode\times\else\math{\times}\fi}
\newcommand\ttt[1]{10\sp{#1}}
\newcommand\tttt[1]{\by\ttt{#1}}

\newbox{\wdbox}
\renewcommand\c{\setbox\wdbox=\hbox{,}\hspace{\wd\wdbox}}
\renewcommand\i{\setbox\wdbox=\hbox{i}\hspace{\wd\wdbox}}

\newcount\timect
\newcount\hourct
\newcount\minct
\newcommand\now{\timect=\time \divide\timect by 60
         \hourct=\timect \multiply\hourct by 60
         \minct=\time \advance\minct by -\hourct
         \number\timect:\ifnum \minct < 10 0\fi\number\minct}

\newcommand\mctc{\multicolumn{2}{c}}

\catcode`@=11

\newcommand\comment[1]{}
\newcommand\commenton{\catcode`\%=14}
\newcommand\commentoff{\catcode`\%=12}

\renewcommand\math[1]{$#1$}
\newcommand\mathshifton{\catcode`\$=3}
\newcommand\mathshiftoff{\catcode`\$=12}

\comment{the backslash is necessary}

\comment{alignment tab}

\let\atab=&
\newcommand\atabon{\catcode`\&=4}
\newcommand\ataboff{\catcode`\&=12}

\let\oldmsp=\sp
\let\oldmsb=\sb
\def\sp#1{\ifmmode
           \oldmsp{#1}%
         \else\strut\raise.85ex\hbox{\scriptsize #1}\fi}
\def\sb#1{\ifmmode
           \oldmsb{#1}%
         \else\strut\raise-.54ex\hbox{\scriptsize #1}\fi}
\newbox\@sp
\newbox\@sb
\def\sbp#1#2{\ifmmode%
           \oldmsb{#1}\oldmsp{#2}%
         \else
           \setbox\@sb=\hbox{\sb{#1}}%
           \setbox\@sp=\hbox{\sp{#2}}%
           \rlap{\copy\@sb}\copy\@sp
           \ifdim \wd\@sb >\wd\@sp
             \hskip -\wd\@sp \hskip \wd\@sb
           \fi
        \fi}
\def\msp#1{\ifmmode
           \oldmsp{#1}
         \else \math{\oldmsp{#1}}\fi}
\def\msb#1{\ifmmode
           \oldmsb{#1}
         \else \math{\oldmsb{#1}}\fi}
\def\supon{\catcode`\^=7}
\def\supoff{\catcode`\^=12}
\def\subon{\catcode`\_=8}
\def\suboff{\catcode`\_=12}
\def\supsubon{\supon \subon}
\def\supsuboff{\supoff \suboff}

\newcommand\actcharon{\catcode`\~=13}
\newcommand\actcharoff{\catcode`\~=12}

\newcommand\paramon{\catcode`\#=6}
\newcommand\paramoff{\catcode`\#=12}

\comment{And now to turn us totally on and off...}

\newcommand\reservedcharson{\commenton \mathshifton \atabon \supsubon \actcharon
	\paramon}

\newcommand\reservedcharsoff{\commentoff \mathshiftoff \ataboff
	\supsuboff \actcharoff \paramoff}

\catcode`@=12
\reservedcharsoff

\reservedcharson

\comment{ Must have ONLY ONE of these... trust these macros, they work

}

\newcommand{\squishlist}{
 \begin{list}{$\bullet$}
  { \setlength{\itemsep}{1pt}
     \setlength{\parsep}{0pt}
     \setlength{\topsep}{3pt}
     \setlength{\partopsep}{0pt}
     \setlength{\leftmargin}{2.0em}
     \setlength{\labelwidth}{1.5em}
     \setlength{\labelsep}{0.5em} } }

\newcommand{\squishend}{
  \end{list}  }

\reservedcharson
\actcharon

\usepackage{ulem}
\shorttitle{Deciphering the Atmospheric Composition of WASP-12b}
\shortauthors{Stevenson {\em et al.}}

\submitted{}

\begin{document}

\title{Deciphering the Atmospheric Composition of WASP-12\lowercase{b}: \\A Comprehensive Analysis of its Dayside Emission}

\author{Kevin B.\ Stevenson\altaffilmark{1}}
\author{Jacob L.\ Bean\altaffilmark{1}}
\author{Nikku Madhusudhan\altaffilmark{2,3}}
\author{Joseph Harrington\altaffilmark{4}}
\affil{\sp{1}Department of Astronomy and Astrophysics, University of Chicago, 5640 S Ellis Ave, Chicago, IL 60637, USA}
\affil{\sp{2}Institute of Astronomy, University of Cambridge, Madingley Road, Cambridge, CB3 0HA, UK}
\affil{\sp{3}Department of Physics \& Department of Astronomy, Yale University, P.O. Box 208120, New Haven, CT 06520, USA}
\affil{\sp{4}Planetary Sciences Group, Department of Physics, University of Central Florida, Orlando, FL 32816-2385, USA}

\email{E-mail: kbs@uchicago.edu}

\begin{abstract}
WASP-12b was the first planet reported to have a carbon-to-oxygen ratio (C/O) greater 
than one in its dayside atmosphere. However, recent work to further characterize its 
atmosphere and confirm its composition has led to incompatible measurements and 
divergent conclusions.  Additionally, the recent discovery of stellar binary companions 
\sim1{\arcsec} from WASP-12 further complicates the analyses and subsequent 
interpretations.  We present a uniform analysis of all available {\em Hubble} and 
{\em Spitzer Space Telescope} secondary-eclipse data, including previously-unpublished 
{\em Spitzer} measurements at 3.6 and 4.5 {\microns}.  The primary controversy in the 
literature has centered on the value and interpretation of the eclipse depth at 4.5 
{\microns}.  Our new measurements and analyses confirm the shallow eclipse depth in 
this channel, as first reported by Campo and collaborators and used by Madhusudhan 
and collaborators to infer a carbon-rich composition.  To explain WASP-12b's observed 
dayside emission spectrum, we implemented several recent retrieval approaches. We 
find that when we exclude absorption due to C\sb{2}H\sb{2} and HCN, which are not 
universally considered in the literature, our models require implausibly large 
atmospheric CO\sb{2} abundances, regardless of the C/O.  By including C\sb{2}H\sb{2} 
and HCN in our models, we find that a physically-plausible carbon-rich solution 
achieves the best fit to the available photometric and spectroscopic data.  In 
comparison, the best-fit oxygen-rich models have abundances that are inconsistent 
with the chemical equilibrium expectations for hydrogen-dominated atmospheres and 
are 670 times less probable.  Our best-fit solution is also 7.3\tttt{6} times more 
probable than an isothermal blackbody model.
\end{abstract}

\section{Introduction}
\label{intro}

The study of exoplanetary atmospheres has shown that planets are a
diverse group of objects and that placing constraints on their
composition and chemistry will advance our understanding of planet
formation and planetary physics.  Detailed characterization of hot
Jupiters is possible when these planets pass in front of or behind
their parent stars.  The latter event, known as the secondary eclipse,
reveals a planet's dayside emission spectrum using measurements at
multiple infrared wavelengths.  By comparing atmospheric models to the
measured spectrum, we can place constraints on the absolute chemical
abundances and thermal profile.

At the time of its discovery, WASP-12b was the most heavily-irradiated
exoplanet yet known, with an equilibrium temperature in excess of 2500
K \citep{Hebb2009}.  This afforded an excellent
opportunity to measure the planet's dayside thermal emission over a
broad range of infrared wavelengths.  These data were used to place
constraints on the planet's atmospheric composition and thermal
profile; however, independent interpretations of the individual data
sets have led to different conclusions.  Therefore, we conducted a
uniform analysis of all available {\em Hubble} and {\em Spitzer Space
  Telescope} secondary eclipse data, including previously-unpublished
{\em Spitzer}\/ measurements at 3.6 and 4.5 {\microns}, to assemble a more consistent
description of the planet's atmospheric composition and thermal
profile.

In a previous report, we used {\em Spitzer} to measure the dayside emission of
WASP-12b at four infrared wavelengths \citep{Campo2011}.  We combined these data with
secondary-eclipse depths measured in the J, H, and Ks bands
\citep{Croll2011} and found that the best-fit atmospheric models
favored a carbon-to-oxygen ratio (C/O) $\ge$ 1 \citep{Madhu2011-wasp12b}.  For 
comparison, the solar C/O is \sim0.54.

Due to WASP-12b's small semi-major axis and inflated radius, the
planet's shape may not be spherical, but that of a prolate spheroid instead.
Using full-orbit observations of WASP-12b with {\em Spitzer},
\citet{Cowan2012} measured significant ellipsoidal variations at 4.5
{\microns}, but no variations at 3.6 {\microns}.  Under this
scenario, they reported eclipse depths that are consistent with
previous results.  However, by fixing the ellipsoidal variations
to zero (the null hypothesis), \citet{Cowan2012} noted that the
measured eclipse depths favor a solar C/O and a modest thermal
inversion.  To make this determination, they varied the abundance of 
CO as a proxy for varying the C/O in their 1D radiative transfer models.

Further obfuscating the planet's atmospheric composition,
\citet{Bergfors2013} announced the discovery of a companion star only
1{\arcsec} (less than one {\em Spitzer}\/ pixel) from WASP-12.  
\citet{Bechter2014} and \citet{Sing2013} have since demonstrated 
that the companion is a binary (labeled WASP-12BC) that is physically 
associated with the primary star WASP-12A.
Upon determining that the companions are of stellar type M0 -- M1,
\citet{Crossfield2012} combined results from a narrow-band,
2.315-{\micron} secondary-eclipse measurement with a corrected,
weighted average of previously-reported eclipse depths, assuming the
null hypothesis from \citet{Cowan2012}.  Following \citet{Barman2001, 
Barman2005}, they also constructed a variety of atmospheric models for 
comparison.  Using $\chi^2$ and Bayesian Information Criterion (BIC) 
values as their metrics, they concluded that a blackbody approximates 
WASP-12b's emission spectrum well, and that its photosphere is nearly 
isothermal.

Both \citet{LopezMorales2010} and \citet{Fohring2013} observed
WASP-12b in the z$^{\prime}$ band (centered at 0.9 {\microns}) during 
secondary eclipse; however, their reported depths (0.082 {\pm} 0.015\% 
and 0.130 {\pm} 0.013\%, respectively) are discrepant by $>3\sigma$.  
This difference may be the result of temporal variability in the planet 
flux or unmodeled systematics in one or both analyses.

\begin{table*}[tb]
\centering
\caption{\label{tab:ObsDates} 
Observation Information}
\begin{tabular}{crccclcc}
    \hline
    \hline
    Label\tablenotemark{a}       
                & Observation Date      & Duration  & Frame Time    & Total Frames  & {\em Spitzer} & Wavelength    & Previous      \\
                &                       & [minutes] & [seconds]     &               & Pipeline      & [\microns]    & Publications\tablenotemark{b}  \\
    \hline
    wa012bs21   &   October 29, 2008    & 338       & 12            & 1560          & S18.25.0       & 4.5           & Ca11, M11     \\
    wa012bs41   &   October 29, 2008    & 338       & 12            & 1560          & S18.25.0       & 8.0           & Ca11, M11     \\
    wa012bs11   &   November 3, 2008    & 367       & 12            & 1697          & S18.25.0       & 3.6           & Ca11, M11     \\
    wa012bs31   &   November 3, 2008    & 367       & 12            & 1697          & S18.25.0       & 5.8           & Ca11, M11     \\
    wa012bs22   &        May 3, 2010    & 460       & 12            & 2109          & S18.18.0       & 4.5           & -             \\
    wa012bs12   &        May 4, 2010    & 460       & 12            & 2109          & S18.18.0       & 3.6           & -             \\
    wa012bs13   &  November 18, 2010    & 427       & 0.4           & 12728         & S18.18.0       & 3.6           & Co12          \\
    wa012bs23   &  December 11, 2010    & 673       & 0.4           & 20000         & S18.18.0       & 4.5           & Co12          \\
    wa012bs24   &  December 12, 2010    & 427       & 0.4           & 12728         & S18.18.0       & 4.5           & Co12          \\
    \hline
\end{tabular}
\tablenotetext{1}{wa012b designates the planet, $s$ specifies
  secondary eclipse, and \#\# identifies the wavelength and
  observation number.}
\tablenotetext{2}{Ca11 = \citet{Campo2011}, M11 = \citet{Madhu2011-wasp12b}, and Co12 = \citet{Cowan2012}.}
\end{table*}

Using observations from the Wide-Field Camera 3 (WFC3) instrument on board the 
{\em Hubble Space Telescope}\/ (HST) and two different atmospheric modeling 
approaches (optimal estimation and Markov-chain Monte Carlo (MCMC) retrieval), 
\citet{Swain2013} found that the companion-stars-corrected
dayside spectrum is best fit by an H\sb{2} atmosphere with no
additional opacity sources.  Such a model supports the isothermal
findings of \citet{Crossfield2012}.  When including the standard
opacity sources (H\sb{2}O, CH\sb{4}, CO, and CO\sb{2}),
\citet{Swain2013} found no evidence for a C/O $\ge$ 1 or a thermal inversion.

Using the published results from \citet{Crossfield2012} and
\citet{Swain2013}, \citet{Line2014-C/O} carried out a temperature and
abundance retrieval analysis of eight exoplanets, including WASP-12b.  
They used a suite of inverse modeling algorithms, called CHIMERA, which employ multiple 
Bayesian retrieval approaches and found two possible atmospheric scenarios.
Their preferred mode (``null'') favors a weak thermal inversion and a large CO\sb{2} 
abundance.  A secondary mode (``ellipsoidal'') results in a slightly stronger thermal 
inversion and an even higher CO\sb{2} mixing ratio.  Both scenarios favor a solar C/O, 
but larger ratios closer to unity cannot be ruled out.

In this paper, we present new broadband secondary-eclipse observations of WASP-12b at 
3.6 and 4.5 {\microns} using {\em  Spitzer}.  We combine these data with reanalyses of 
previously published {\em Spitzer} InfraRed Array Camera \citep[IRAC, ][]{IRAC} eclipse 
observations \citep{Campo2011,Cowan2012} and emission-spectroscopy observations using 
{\em HST}/WFC3\/ \citep{Swain2013}.  We also account for the contamination by WASP-12BC. 
The work presented here tests a variety of modeling approaches, including that of 
\citet{Line2014-C/O}, and offers a comprehensive and uniform analysis of available 
WASP-12b secondary-eclipse data to constrain its dayside atmospheric composition.

\section{SPITZER/IRAC OBSERVATIONS AND DATA ANALYSIS}
\label{sec:iracobs}

\subsection{Observations and Reduction}

For the new observations presented here, {\em Spitzer}'s IRAC acquired 2109 frames 
of WASP-12 in each of the 3.6 and 4.5 {\micron} channels (Program 60003, PI
Joseph Harrington).  As with the observations presented by
\citet{Campo2011}, we used 12-second exposures in full-frame mode and
achieved a duty cycle of almost 80\%.  Conversely, the observations
presented by \citet[][Program 70060, PI Pavel
 Machalek]{Cowan2012} used 0.4-second exposures in subarray mode and,
due to a 104-second gap between subarray sets, have a
duty cycle of $\sim18\%$.  Therefore, the latter
achieved approximately half of the precision obtained with the
full-array observations.  Additional observation information is listed
in Table \ref{tab:ObsDates}.

We produce systematics-corrected light curves using the Photometry for
Orbits, Eclipses, and Transits (POET) pipeline
\citep{Campo2011,Stevenson2011,Cubillos2013}.  POET flags bad pixels using a
two-iteration, 4$\sigma$ filter along the time axis of each set of 64 frames, 
calculates image centers from a 2D Gaussian fit, and applies 5$\times$ 
interpolated aperture photometry \citep{Harrington2007} for apertures up to 
5.0 pixels in radius.  It then removes systematics and fits lightcurve models 
as described below.

\subsection{Light-Curve Systematics and Fits}

{\em Spitzer} light curves exhibit several well characterized
systematics \citep{Charbonneau2005, Agol2010, Knutson2011,
  Stevenson2011, Lewis2013}.  We test polynomial and exponential
functions when modeling the time-dependent systematics at all wavelengths and apply
Bilinearly-Interpolated Subpixel Sensitivity (BLISS) mapping
\citep{Stevenson2011} to model the position-dependent systematics
at 3.6 and 4.5 {\microns}, except for wa012bs21, which does not exhibit
this effect.  This is unusual for this array, but has been seen occasionally 
\citep[e.g., ][]{Todorov2010}.

\begin{figure}[tb]
\centering
\includegraphics[width=1.0\linewidth,clip]{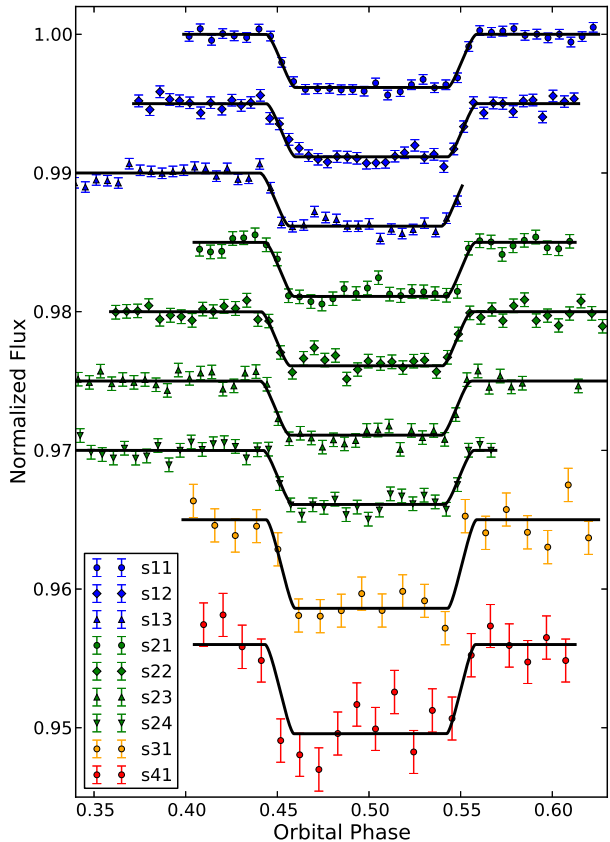}
\caption{\label{fig:spitzerlc}{
WASP-12b photometric light curves using {\em Spitzer}/IRAC.  The
results are corrected for systematics, normalized to the system flux,
and shifted vertically for ease of comparison.  The lines are best-fit
models and the error bars are 1\math{\sigma} uncertainties.  The
shorthanded legend labels correspond to the last three characters in
each event's label (e.g., s11 = wa012bs11).
}}
\end{figure}

Simultaneously with the systematics, we fit the secondary eclipses
with the uniform-source equations from 
\citet{mandel2002}.  We perform a joint, simultaneous fit of the 3.6
and 5.8 {\micron} observations as well as a separate joint fit for the
4.5 and 8.0 {\micron} observations.  For the 2008 observations, IRAC
observed these channel pairs simultaneously, so the fits share the
eclipse midpoint.  In each joint fit, the light curves at 
the same wavelength share one eclipse depth.  We estimate uncertainties using
two techniques, differential-evolution MCMC (DE-MCMC) and
residual permutation.  The latter produces slightly larger
uncertainties, which we adopt.  Using the transit parameters from
\citet{Stevenson2013}, we fix the eclipse duration (0.11459 orbits)
and ingress/egress times (0.01557 orbits).  Allowing these parameters
to vary does not change our final results.  Figure \ref{fig:spitzerlc}
displays binned, systematics-corrected light curves with best-fit
models.

Individual analyses of the {\em Spitzer}\/ eclipses produced depths that 
are consistent with the joint fits to within 1$\sigma$ at 3.6 {\microns} 
and to within 2$\sigma$ at 4.5 {\microns}.  See Table \ref{tab:EclDepths} 
for the individual eclipse depths using a 3.0 pixel aperture size.  It is 
intriguing that both 4.5 {\micron} eclipse depth measurements extracted from 
the phase curve observation are deeper than the remaining secondary eclipse 
measurements.

As with \citet{Cowan2012}, we do not include the 3.6-{\micron}
secondary eclipse from 2010 November 17 in our final analysis.  This
is due to the presence of a strong feature (possibly due to stellar
activity) during the latter half of the eclipse that alters the
measured depth.  In contrast to \citet{Cowan2012}, we do not fit the
entire phase curves when determining the eclipse depths.  This is to
ensure that unmodeled flux variations in the phase curves do not
affect the measured depths and bypasses the question of ellipsoidal 
variation.  As reported by \citet{Stevenson2013},
when we do fit the full phase curves, our best-fit models confirm the
large ellipsoidal variations in only the 4.5 {\micron} channel.  Our
measured eclipse depths are in excellent agreement ($<1\sigma$)
with those favored by \citet[][ellipsoidal variation models]{Cowan2012} 
and are inconsistent with the null hypothesis (no ellipsoidal variation) 
4.5 {\micron} depth by $10\sigma$.
\\

\begin{table}[tb]
\centering
\caption{\label{tab:EclDepths} 
Individual Eclipse Depths Using a 3.0 Pixel Aperture Size}
\begin{tabular}{cc}
    \hline
    \hline      
    Label       & Eclipse Depth     \\
                & [\%]              \\
    \hline
    wa012bs11   & 0.41 {\pm} 0.02   \\
    wa012bs12   & 0.38 {\pm} 0.02   \\
    wa012bs13   & 0.36 {\pm} 0.02   \\
    wa012bs21   & 0.38 {\pm} 0.02   \\
    wa012bs22   & 0.36 {\pm} 0.02   \\
    wa012bs23   & 0.42 {\pm} 0.02   \\
    wa012bs24   & 0.42 {\pm} 0.02   \\
    \hline
\end{tabular}
\end{table}

\subsection{Dilution Factor Correction}

\begin{figure}[tb]
\centering
\includegraphics[width=1.0\linewidth,clip]{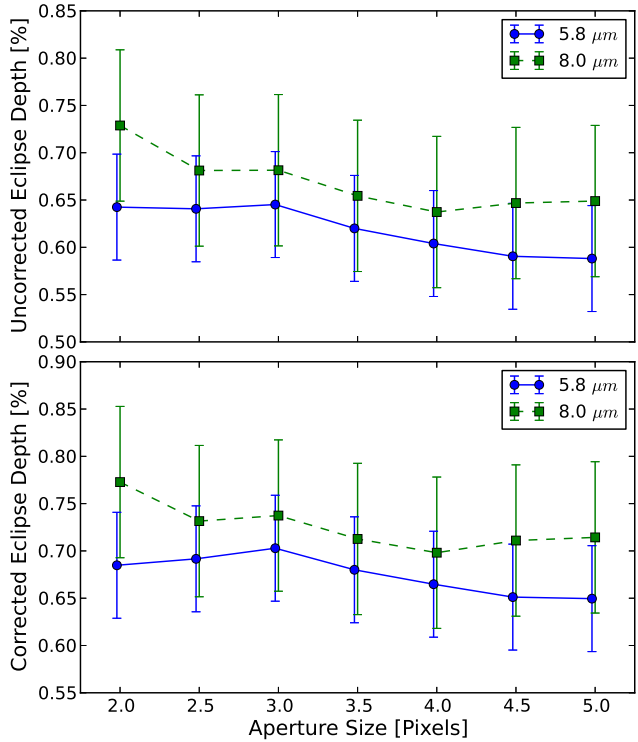}
\caption{\label{fig:apsizes}{
Measured (top) and companion-star corrected (bottom) eclipse depths at
5.8 and 8.0 {\microns}.  The measured eclipse depths may have a weak
dependence on photometry aperture size; however, this trend is not
significant as the points all fall within 1$\sigma$ of each other.
There is no discernible trend at 3.6 or 4.5 {\microns}.
}}
\end{figure}

A recently discovered, binary companion
\citep{Bergfors2013,Bechter2014} resides well within the {\em Spitzer}
photometry apertures, thus diluting the measured eclipse depths.  To
correct for this effect, we apply the dilution factors calculated by
\citet[$\alpha_{Comp}(3.6,4.5,5.8,8.0)$ = 0.1149, 0.1196, 0.1207,
  0.1190]{Stevenson2013} to each of the four {\em Spitzer} channels
using the equation:
\begin{equation}
\label{eqn:ecldepth}
\delta\sb{Corr}(\lambda) = \left [1+g(\beta,\lambda)\alpha_{Comp}(\lambda)\right]\delta\sb{Meas}(\lambda),
\end{equation}
\noindent where $\delta\sb{Meas}(\lambda)$ are the measured (or
uncorrected) eclipse depths and $g(\beta,\lambda)$ are the
wavelength-dependent companion flux fractions inside a photometric
aperture of size $\beta$.  Table \ref{tab:depths} gives the final
eclipse depths.  Since we apply a single eclipse depth to fit all of
the observations from a given channel, we select a single aperture size
for each channel, thus allowing us to apply a single
$g(\beta,\lambda)$ value during the correction.  In our final
analysis, we use an aperture size of 3.0 pixels for all channels.  We
tested aperture sizes up to a radius of 5.0 pixels in all channels and
found no significant (>1$\sigma$) correlation with the measured
eclipse depths.  See Figure \ref{fig:apsizes} for examples at 5.8 and
8.0 {\microns}.

\begin{table}[tb]
\centering
\caption{\label{tab:depths} 
Companion-Star-Corrected Eclipse Depths}
\begin{tabular}{ccr@{\,{\pm}\,}l}
    \hline
    \hline      
    Wavelength      & RMS   & \mctc{Eclipse Depth}    \\
    ({\microns})    & (ppm) & \mctc{(\%)}             \\
    \hline
    1.10 -- 1.15    & 1512   & 0.119   & 0.017        \\
    1.15 -- 1.20    & 1374   & 0.128   & 0.012        \\
    1.20 -- 1.25    & 1263   & 0.101   & 0.012        \\
    1.25 -- 1.30    & 1203   & 0.142   & 0.011        \\
    1.30 -- 1.35    & 1274   & 0.154   & 0.012        \\
    1.35 -- 1.40    & 1242   & 0.156   & 0.012        \\
    1.40 -- 1.45    & 1296   & 0.184   & 0.012        \\
    1.45 -- 1.50    & 1299   & 0.198   & 0.012        \\
    1.50 -- 1.55    & 1407   & 0.196   & 0.013        \\
    1.55 -- 1.60    & 1563   & 0.179   & 0.014        \\
    1.60 -- 1.65    & 1760   & 0.192   & 0.017        \\
    \hline
    3.6             & 2296   & 0.421    & 0.011       \\
                    & 2523                            \\
                    & 6326                            \\
    4.5             & 3214   & 0.428    & 0.012       \\
                    & 3131                            \\
                    & 8182                            \\
                    & 8251                            \\
    5.8             & 10633 & 0.696    & 0.060        \\
    8.0             & 13240 & 0.696    & 0.096        \\
    \hline
\end{tabular}
\end{table}

\section{HST/WFC3 OBSERVATIONS AND DATA ANALYSIS}
\label{sec:wfc3obs}

\subsection{Observation and Reduction}

Spanning five orbits on 2011 April 15, {\em HST} observed a
secondary eclipse of WASP-12b using the WFC3 instrument with its G141
grism.  \citet{Swain2013} provide additional details on the
observations (Program 12230, PI Mark Swain).  Using the
reduction, extraction, and calibration steps described by
\citet{Stevenson2013}, we generate eleven wavelength-dependent light
curves spanning 1.10 -- 1.65 {\microns}.  See \citet{Berta2012}, 
\citet{Deming2013}, \citet{Sing2013}, and \citet{Kreidberg2014} 
for additional discussion on WFC3 analyses and calibration.

\subsection{Light-Curve Systematics and Fits}
\label{sec:hst-nasc}

\begin{figure}[tb]
\centering
\includegraphics[width=1.0\linewidth,clip]{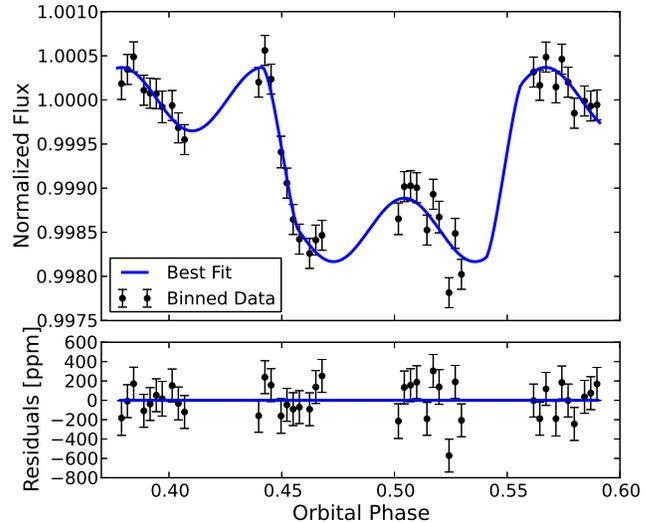}
\caption{\label{fig-wfc3white}{
WASP-12b band-integrated light curve (top panel) with residuals
(bottom panel) from 2011 April 15 using {\em HST}'s WFC3 instrument.
The black data points are binned in phase and display 1\math{\sigma}
uncertainties.  The solid blue line depicts the best-fit model, which
includes a sinusoidal function to model the effects of telescope
thermal breathing.
}}
\end{figure}

These data do not exhibit the strong persistence behavior between
buffer dumps that is seen in some other WFC3 exoplanet light curves.
We do, however, detect evidence for light-curve
fluctuations due to thermal breathing of the telescope as it warms and
cools while orbiting the Earth every $\sim$96 minutes (see Figure
\ref{fig-wfc3white}).  Previous analyses detect similar variations in
the WFC3 WASP-12b transmission spectroscopy observations
\citep{Sing2013,Stevenson2013}.
To model the white light curve, we use the uniform-source equations
from \citet{mandel2002} for the secondary eclipse over orbits 2 -- 5,
a linear slope for the baseline, and a sinusoidal function for the
thermal breathing.

\begin{figure}[tb]
\centering
\includegraphics[width=1.0\linewidth,clip]{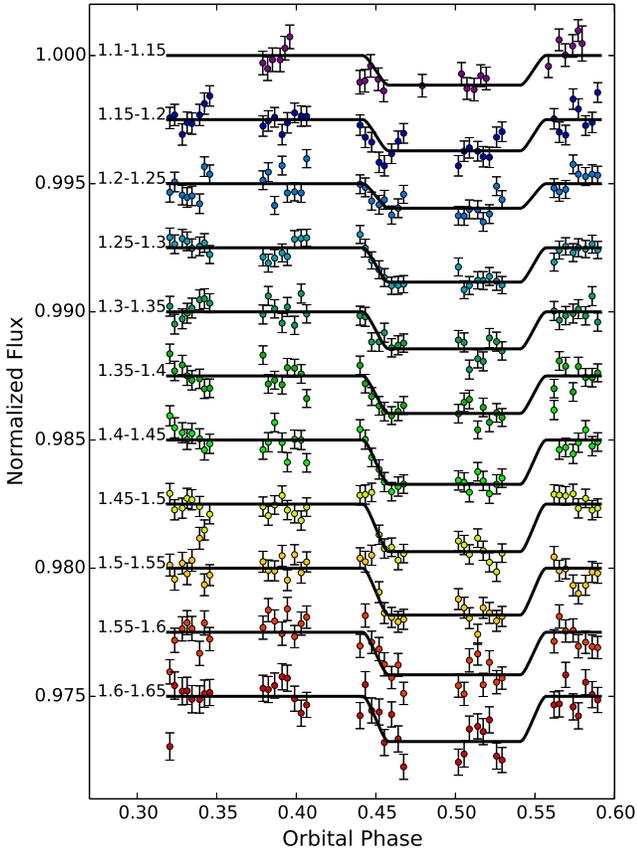}
\caption{\label{fig:hstlc}{
WASP-12b spectroscopic light curves from 2011 April 15 using {\em 
HST}'s WFC3 instrument.  The {\tt Divide-White} method
\citep{Stevenson2013} produced these results, which are binned,
normalized to the system flux, and shifted vertically for ease of comparison.  
The error bars are 1\math{\sigma} uncertainties and the
black lines are best-fit models.  The wavelength range for each of the
11 channels is specified in {\microns}.  For the bluest channel, we do
not model the first orbit or the final batch within each orbit because
the flux is systematically higher than the other batches.
}}
\end{figure}

To model the spectroscopic light curves, we apply both methods
described by \citet{Stevenson2013}.  Method 1 uses the same functional
form as the white light-curve analysis, with wavelength-independent
systematic models and wavelength-dependent eclipse depths and baseline offsets.
Method 2, also called {\tt Divide-White}, fits all of the orbits using
the white light curve to generate a non-analytic model of the
wavelength-independent systematics.  The only free parameters with
this model are the wavelength-dependent secondary eclipse depths and
baseline offsets.  We estimate uncertainties with our DE-MCMC
algorithm. In agreement with \citet{Swain2013}, correlation plots of
RMS {\vs} bin size indicate that there is no significant
time-correlated noise in the data and, as such, there is no need to
inflate uncertainty estimates \citep{Pont2006,Winn2008b}.  
The WFC3 dataset has an insufficient number of points for a residual-permutation 
analysis.  We plot the normalized spectroscopic light curves from Method 2 in
Figure \ref{fig:hstlc}.  The residual RMS values range from 1190 to
1640 ppm and the uncertainties range from 1.07 to 1.28$\times$ the
photon limit, with an average of 1.15$\times$.

\subsection{Dilution Factor Correction}

\begin{figure}[tb]
\centering
\includegraphics[width=1.0\linewidth,clip]{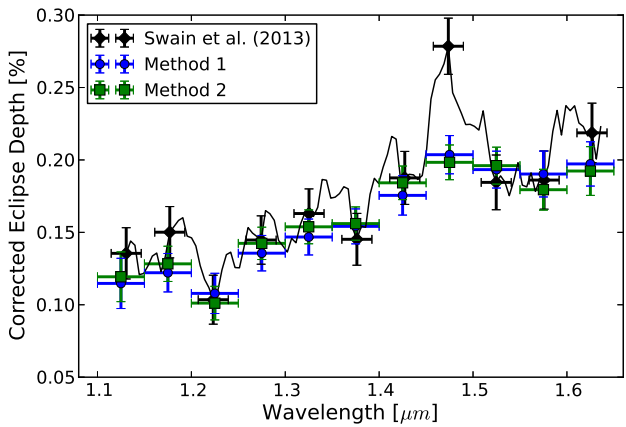}
\caption{\label{fig:hstspec}{
WASP-12b corrected emission spectrum using WFC3's G141 grism.  Both
methods used in our analyses (blue circles and green squares) agree
with the results from \citet[][black line with diamonds for
comparison]{Swain2013} in all but one of the spectroscopic channels.
}}
\end{figure}

The spectroscopic extraction technique employed above does not
separate the WASP-12 signal from that of the companion stars.
Therefore, we estimated the corrected eclipse depths in Table
\ref{tab:depths} using Equation \ref{eqn:ecldepth} (where
$g(\beta,\lambda)=1$) and the companion star dilution factors given in
Table 4 of \citet{Stevenson2013}.  Figure \ref{fig:hstspec} displays
the corrected eclipse depths from both techniques and compares the
results to those from \citet{Swain2013}.  All but one of the
spectroscopic channels agree to within 1$\sigma$.  The source of the
outlier is unknown.  We apply Method 2 for the remainder of our
analysis.

\section{ATMOSPHERIC MODELS AND DISCUSSION}
\label{sec:atm}

When deriving the best-fit atmospheric models, we use the eleven
spectroscopic and four photometric eclipse depths listed in Table
\ref{tab:depths}.  Additionally, we use the four ground-based secondary-eclipse depths
published by \citet{LopezMorales2010} and \citet{Croll2011}, after
correcting for the contribution from the companion stars.  We find
corrected depths of 0.085 {\pm} 0.016\%, 0.140 {\pm} 0.030\%, 0.191
{\pm} 0.020\%, and 0.340 {\pm} 0.014\% in the z\sp{$\prime$}, J, H,
and K bands, respectively.  Despite attempts to include additional 
photometric measurements by \citet{Fohring2013} and \citet{Crossfield2012}, 
their reported eclipse depths are inconsistent with all of our atmospheric 
models.  \citet{Fohring2013}'s suggestion of variability may be unlikely given the 
observed consistency in measured {\em Spitzer} eclipse depths (see Table
\ref{tab:EclDepths}).  We recommend that additional observations with longer 
out-of-eclipse baselines be acquired in these bandpasses to establish more precise and 
consistent eclipse depths.  For completeness, we discuss below how these 
measurements compare to our derived models.

Using the observed dayside emission spectrum, we apply the atmospheric
modeling and retrieval technique described by \citet{Madhu2012} to
place constraints on the properties of WASP-12b's atmosphere.  We
compute model spectra using 1D line-by-line radiative transfer in a
plane-parallel atmosphere.  This approach assumes local thermodynamic
equilibrium, hydrostatic equilibrium, and global energy balance.
The models make no assumption about the layer-by-layer radiative equilibrium 
and, as with the models of \citet{Line2014-C/O}, impose no constraint on the 
atmospheric chemical abundances.  Thus, our atmospheric retrievals explore 
both physically plausible and implausible regions of the parameter space.

In this work, we consider three sets of model atmospheres.  The first includes 
line-by-line molecular absorption due to H$_2$O, CO, CH$_4$, and CO$_2$; the 
second also considers absorption due to C$_2$H$_2$ and HCN 
\citep[see,  e.g., ][]{Madhu2012, Moses2013}.  Both sets include H\sb{2}-H\sb{2} 
collision-induced opacities and assign six free parameters for the 
pressure-temperature profile.  TiO and VO do not have features in the wavelength 
region sampled by these data and are not included.  The third set is an isothermal 
blackbody model that has only one free parameter, the temperature.

When considering only the four primary molecular absorbers, we find a bimodal 
distribution in the C/O.  The C-rich mode (C/O $\gtrsim$ 1) achieves a better fit 
than the O-rich mode (C/O $\sim$ 0.5, $\Delta$BIC = 9.5, $\sim$120 times more 
probable); however, both modes require physically implausible atmospheric abundances.  
Specifically, the best-fit model requires high CH\sb{4} and CO\sb{2} abundances 
(4.3\tttt{-3} and 9.9\tttt{-5}), with very little H\sb{2}O and CO (4.8\tttt{-8} and 
9.1\tttt{-10}, respectively).  However, \citet{Madhu2012} and \citet{Moses2013} 
demonstrate that the CO\sb{2} abundance in a hot, hydrogen-dominated atmosphere cannot 
exceed that of H\sb{2}O or CO.  The solution to this problem lies in the addition of 
C\sb{2}H\sb{2} and HCN to our atmospheric models.

With six molecular absorbers, we explore both O- and C-rich scenarios.  For the former, 
we would expect to detect a broad H\sb{2}O absorption feature in the {\em HST}/WFC3 
spectroscopic data.  This is not the case, so the O-rich models must adopt a 
predominantly isothermal profile at pressures $\gtrsim 0.01$ bar (which are the depths 
probed by WFC3) and decrease the H\sb{2}O abundance by a factor of five relative to 
solar composition.  Furthermore, C\sb{2}H\sb{2} and HCN are not thermochemically favored 
in an O-rich atmosphere; therefore, to fit the shallow eclipse depth at 4.5 {\microns}, 
the models compensate by increasing the CO\sb{2} abundance by two orders of magnitude 
relative to solar composition.  As a result, the best-fit O-rich model ($\chi^2 \sim 50$) 
remains physically implausible with its strong CO\sb{2} feature at 4.5 {\microns} and 
insignificant H\sb{2}O absorption.  The lack of H\sb{2}O absorption in the WFC3 bandpass 
and the low 4.5 {\micron} photometry point are more readily explained by C-rich models 
($\chi^2 \sim 38$).  These models naturally explain the lack of H\sb{2}O due to 
insufficient oxygen after the formation of CO, and they utilize C\sb{2}H\sb{2} and HCN 
(in addition to CO\sb{2}) to explain the absorption at 4.5 {\microns}.

Although the WFC3 measurements are consistent with an isothermal blackbody model, the 
broadband {\em Spitzer} points preclude such an option.  The 4.5 {\micron} eclipse depth, 
which we derived from four independent datasets with consistent results, is discrepant 
from the isothermal model at a significance of $7\sigma$.

In Table \ref{tab:models}, we compare our best-fit models from the six-molecule and 
isothermal-blackbody scenarios to the photometric and spectroscopic data. Table 
\ref{tab:models} also presents differences in BIC values.  BIC is similar to $\chi^2$, 
but it adds a penalty for using additional free parameters; therefore, smaller BIC values 
are preferable \citep{Liddle2008}.  Using this information, we conclude that the C-rich 
model is 670 times more probable than the O-rich model and7.3\tttt{6} times more probable 
than a blackbody.

\begin{table}[tb]
\centering
\caption{\label{tab:models} 
Atmospheric Model Comparison}
\begin{tabular}{crrrr}
    \hline
    \hline      
    Model Type  & $\chi^2_{\rm Phot}$   & $\chi^2_{\rm Spec}$   & $\chi^2_{\rm Total}$  & $\Delta$BIC   \\
    \hline
Carbon-rich     & 17.2              & 20.4              &  37.6             &   0.0 \\
Oxygen-rich     & 31.0              & 19.6              &  50.6             &  13.0 \\
Blackbody       & 79.9              & 19.1              &  99.0             &  31.6 \\
    \hline
\end{tabular}
\end{table}

\begin{table}[tb]
\centering
\caption{\label{tab:abundances} 
Best-Fit Molecular Abundances}
\begin{tabular}{cccc}
    \hline
    \hline      
    Model        & H\sb{2}O\tablenotemark{a}       
                                  & CO           & CH\sb{4}         \\
    \hline
Oxygen-rich      & 5.0\tttt{-4}   & 5.0\tttt{-4} & 1.0\tttt{-7}     \\
Carbon-rich      & 2.3\tttt{-7}   & 3.4\tttt{-4} & 8.3\tttt{-5}     \\
    \hline
                 & CO\sb{2}       & C\sb{2}H\sb{2} & HCN            \\
    \hline
Oxygen-rich      & 6.7\tttt{-5}   & 1.6\tttt{-10}  & 1.0\tttt{-7}   \\
Carbon-rich      & 9.0\tttt{-7}   & 1.0\tttt{-5}   & 1.0\tttt{-6}   \\
    \hline
\end{tabular}
\tablenotetext{1}{The H\sb{2}O abundance in the C-rich model has very little impact on the observed spectrum at these low levels and can easily be a factor of four larger, thus maintaining physical plausibility.}
\end{table}

Table \ref{tab:abundances} lists the derived molecular abundances for the best-fit, 
six-molecule O- and C-rich models; their carbon-to-oxygen ratios are 0.5 and 1.2, 
respectively.  We compare these results to the best-fit ellipsoidal solution presented 
by \citet[][\{H\sb{2}O, CO, CH\sb{4}, CO\sb{2}\} = \{5.12\tttt{-4}, 2.17\tttt{-3}, 
2\tttt{-10}, 1.07\tttt{-1}\}]{Line2014-C/O}, which uses a 4.5 {\micron} eclipse depth 
that is consistent with our own result.  Other spectroscopic and photometric data points 
from their ellipsoidal solution are also generally consistent with, but not necessarily 
identical to, our own measurements.  For example, our eclipse depths at 3.6 and 4.5 
{\microns} are 4.0 and 2.8 times more precise.  We find that our best-fit C-rich model 
favors $6.4\times$ less CO, $\sim415,000\times$ more CH\sb{4}, and $\sim63,000\times$ 
less CO\sb{2}.  The latter two molecular abundances are outside of the 68\% confidence 
intervals published by \citet{Line2014-C/O}.  Their lack of CH\sb{4} can be explained 
by their preferred low C/O; however, their 10\% CO\sb{2} abundance is irreproducible, 
even when compared to our four-molecule fits, whose implausibly large CO\sb{2} abundances 
do not exceed 1\tttt{-4}.

\citet{Madhu2012} and \citet{Moses2013} demonstrate that in an O-rich, hydrogen-dominated 
atmosphere, the concentration of CH\sb{4}, CO, or CO\sb{2} cannot exceed that of H\sb{2}O, 
regardless of its state of chemical equilibrium.  \citet{Line2014-C/O} list CO and CO\sb{2} 
abundance ratios in their WASP-12b best-fit solutions that far exceed that of H\sb{2}O.  
\citet{Madhu2012} and \citet{Moses2013} also determine that the CO abundance in a hot, 
hydrogen-dominated atmosphere must exceed that of CH\sb{4} and CO\sb{2}.  Again, 
\citet{Line2014-C/O} report a best-fit CO\sb{2} value that is inconsistent with this theory.  
Finally, \citet{Line2014-C/O} do not include C\sb{2}H\sb{2} or HCN in their abundance 
retrieval analysis.  Both molecules are expected to be prevalent in a C-rich atmosphere 
and both have features in {\em   Spitzer}'s 4.5~{\micron} bandpass.  Without these 
molecules, \citet{Line2014-C/O} rely on an unrealistically-large CO\sb{2} abundance to 
explain the relatively shallow eclipse depth at 4.5 {\microns}.  

\begin{figure*}[h!]
\centering
\includegraphics[width=0.91\linewidth,clip]{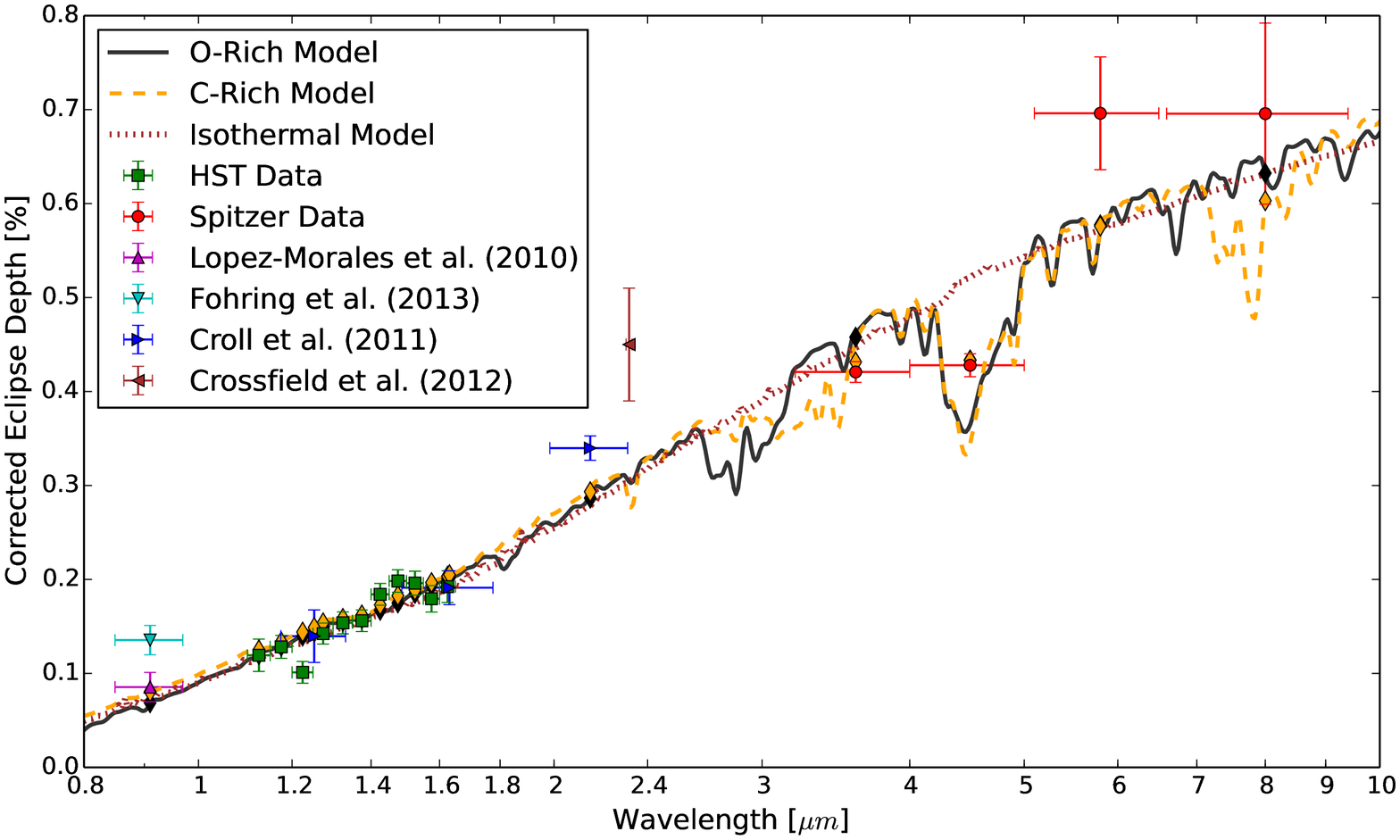}
\caption{\label{fig:corrspec}{
WASP-12b corrected dayside emission spectrum with atmospheric models.
We fit oxygen-rich, carbon-rich, and isothermal blackbody atmospheric
models (solid black, dashed orange, and dotted brown lines,
respectively) to the {\em HST}/WFC3 points (green squares) in the NIR,
the {\em Spitzer}/IRAC points (red circles) from 3 -- 10 {\microns},
and the ground-based points from \citet{LopezMorales2010} and
\citet{Croll2011}.  For reference, we also plot the z\sp{$\prime$} and
2.315-{\micron} measurements from \citet{Fohring2013} and
\citet{Crossfield2012}, respectively.  The isothermal model has a
blackbody temperature of 2930 K.  The best-fit oxygen-rich model requires 
$5\times$ less H\sb{2}O and $\sim100\times$ more CO\sb{2} than solar 
composition.  This physically-implausible scenario achieves a better fit 
than all oxygen-rich, solar-composition models.  However, in comparing 
the bandpass-integrated models (colored diamonds) to the available data, 
the carbon-rich model achieves the best fit by a $\Delta$BIC of 13.0 
(670 times more probable than the best O-rich model).
}}
\end{figure*}

\begin{figure*}[h!]
\centering
\includegraphics[width=0.91\linewidth,clip]{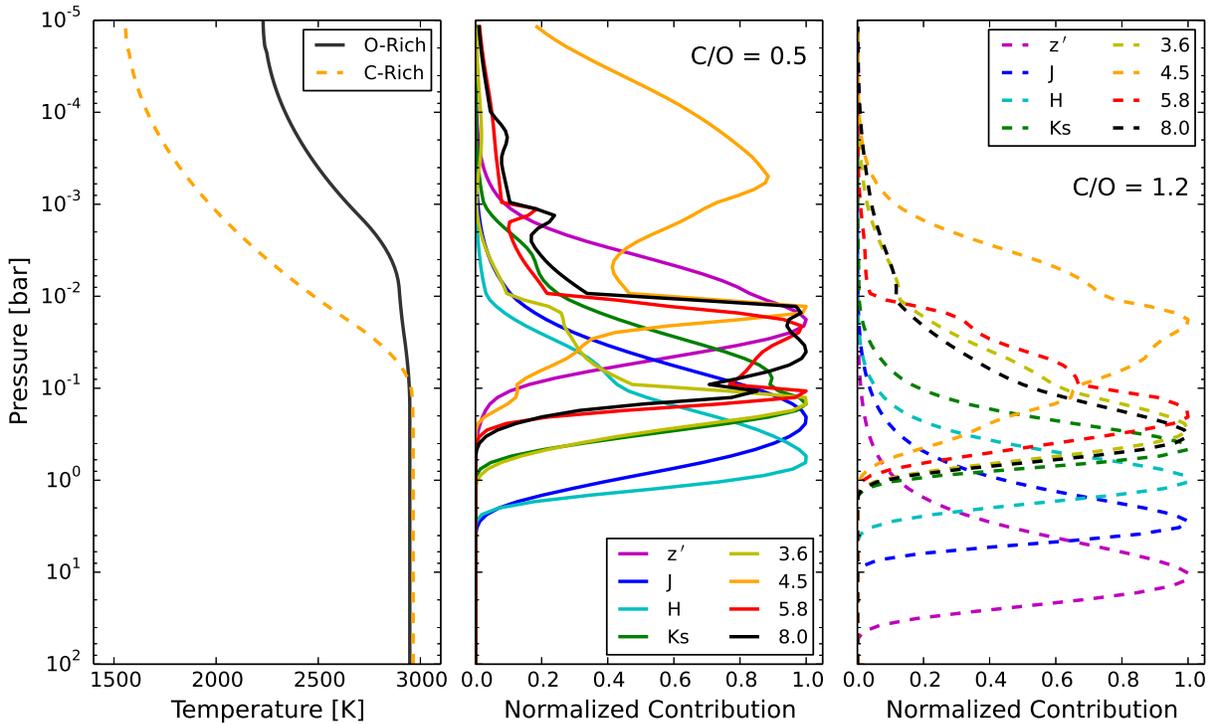}
\caption{\label{fig:pt}{
WASP-12b pressure-temperature profiles and contribution functions.
The left panel shows that the O-rich (solid line) and C-rich (dashed
line) models have monotonically decreasing temperature profiles with 
decreasing pressure.  The center and
right panels illustrate the atmospheric flux origin observed in
each photometric bandpass for the O-rich and C-rich models,
respectively.  The majority of {\em HST}/WFC3's contribution (not
shown) resides in the nearly-isothermal region deeper than 0.01 bar.
}}
\end{figure*}

In Figure \ref{fig:corrspec}, we present the corrected dayside
emission spectrum of WASP-12b and the best-fit atmospheric models 
(which include C\sb{2}H\sb{2} and HCN).  For reference, we 
also add the z\sp{$\prime$} secondary-eclipse measurement from
\citet{Fohring2013} and the narrow-band 2.315-{\micron} 
measurement from \citet{Crossfield2012}.  Although none of the atmospheric
models provides a reasonable fit to these additional data points, the
depth measured by \citet{Fohring2013} further decreases the prospect
of an O-rich atmosphere, while the depth reported by
\citet{Crossfield2012} relies on fitting short baselines before and
after secondary eclipse.  There have been numerous ground-based broadband 
photometry measurements of transiting exoplanets with reported depths in 
excess of model predictions \citep[e.g., ][]{RogersJ2009, Gillon2009, 
Gibson2010, Croll2011}.  \citet{Rogers2013} suggest that, in the event of 
red noise, these measurements may be biased in one direction or another, 
thus making ground-based photometry measurements less reliable than 
previously thought.

In Figure \ref{fig:pt}, we present the thermal profiles and flux contribution 
functions for the O- and C-rich models from Figure \ref{fig:corrspec}.  In 
contrast to profiles presented by \citet{Line2014-C/O}, neither scenario favors 
a thermal inversion.  Our best-fit C-rich thermal profile is constant at pressure 
levels $\gtrsim 0.1$ bar, which is in good agreement with the results presented 
by \citet{Line2014-C/O}; however, the profiles diverge as our temperature 
decreases monotonically with decreasing pressure and their temperature increases, 
thus indicating an inversion.

\citet{Sing2013} and \citet{Stevenson2013} both present evidence for clouds or 
hazes in the atmosphere of WASP-12b at its terminator.  However, light paths 
through the atmosphere are much shorter ($\sim40\times$) with emission spectroscopy 
than they are with transmission spectroscopy, given the latter's slant optical path 
length \citep{Fortney2005}.  Therefore, the presence of clouds or hazes should have 
a smaller cumulative effect on the observed emission spectrum.  The detection of 
spectral features in the dayside emission spectrum rules out the presence of a 
fully opaque, high-altitude dayside cloud layer.  If a thick cloud layer does exist 
on the dayside, it must be at pressure levels $\gtrsim 0.1$ bar, where the thermal 
profile is isothermal.

\section{CONCLUSIONS}
\label{sec:concl}

Through our uniform reanalysis of all available WASP-12b secondary-eclipse data
from both {\em HST}\/ and {\em Spitzer}, we have provided a consistent dataset
from which to draw atmospheric conclusions.  This is particularly important for the 
three 3.6 {\micron} and four 4.5 {\micron} {\em Spitzer} observations, which no
longer exhibit discrepant eclipse depths.  This new analysis also uniformly corrected 
the measured eclipse depths due to contamination from the binary companion WASP-12BC.

To explain WASP-12b's observed dayside emission spectrum, we examined three sets of 
model atmospheres (four molecules, six molecules, and an isothermal blackbody).  All 
models that consider molecular absorption due to only H\sb{2}O, CO, CH\sb{4}, and 
CO\sb{2} require physically implausible atmospheric abundances.  Nonetheless, these 
models find that a C-rich scenario is $\sim$120 times more probable ($\Delta$BIC = 9.5) 
than an O-rich scenario.  With the addition of C\sb{2}H\sb{2} and HCN, the C-rich 
models find physically plausible solutions and continue to achieve the best fits.  The 
combination of photometric and spectroscopic data rule out the best-fit six-molecule 
O-rich and isothermal models at a high statistical significance ($\Delta$BIC = 13.0 and 
31.6, respectively).  A non-isothermal emission spectrum can be confirmed visually in 
Figure \ref{fig:corrspec}, where the isothermal model is rejected at 4.5 {\microns} with 
a significance of 7$\sigma$.  We conclude that, when we account for opacity due to 
C\sb{2}H\sb{2} and HCN, a dayside atmosphere with a C/O $\ge 1$ is the most plausible 
scenario.  We also emphasize that the inclusion of chemical limits in the Bayesian
phase-space exploration has a major effect on the composition retrieval.
These results reaffirm that the C/O is an important facet to consider 
in exoplanet characterization that may provide clues to likely planet 
formation and migration scenarios.

Forthcoming WFC3 data will provide a high-precision correction to WASP-12's stellar 
companion and will place tighter constraints on the planet's transmission spectrum.  
Thus, we leave for future work the application of a temperature and abundance retrieval 
method to both the transmission and emission spectra.  Such work will attempt to 
assemble a more consistent description of WASP-12b's composition, thermal profile, 
and C/O between its dayside and terminator regions.

\acknowledgments

We thank contributors to SciPy, Matplotlib, and the Python Programming
Language, the free and open-source community, the NASA Astrophysics
Data System, and the JPL Solar System Dynamics group for software and
services.  This research made use of Tiny Tim/Spitzer, developed by
John Krist for the Spitzer Science Center. The Center is managed by
the California Institute of Technology under a contract with NASA.
Funding for this work has been provided by NASA grants NNX13AF38G and
NNX13AJ16G. J.L.B. acknowledges support from the Alfred P.~Sloan
Foundation.  N.M. acknowledges support from the Yale Center for Astronomy 
and Astrophysics (YCAA) at Yale University through the YCAA prize
fellowship.
\\

\bibliography{ms}

\end{document}